\documentclass[12pt]{iopart}
\expandafter\let\csname equation*\endcsname\relax
\expandafter\let\csname endequation*\endcsname\relax
\usepackage{amsmath,amsfonts,amssymb,cancel}
\usepackage{graphicx,enumitem}
\usepackage{standalone}
\usepackage{multirow,dcolumn}
\usepackage{txfonts}
\usepackage{hyperref}
\usepackage{soul,cancel}
\usepackage{epsf,amssymb,verbatim,color,multirow,pifont,graphicx,cleveref,tabularx}
\usepackage[dvipsnames]{xcolor}
\usepackage{rotating}

\usepackage{tikz}
\usetikzlibrary{matrix}
\usepackage{braket}

\date{}

\begin{document}

\title[]{Divide-and-conquer embedding for QUBO quantum annealing}

\author{Minjae Jo$^{1}$, Michael Hanks$^{1,\dag}$ and M. S. Kim$^{1}$}
\address{$^{1}$QOLS, Blackett Laboratory, Imperial College London SW7 2AZ, UK}
\ead{$^\dag$m.hanks@imperial.ac.uk}

\vspace{10pt}

\begin{abstract}
Quantum annealing promises to be an effective heuristic for complex NP-hard problems. However, clear demonstrations of quantum advantage are wanting, primarily constrained by the difficulty of embedding the problem into the quantum hardware. Community detection methods such as the Girvin--Newman algorithm can provide a divide-and-conquer approach to large problems. Here, we propose a problem-focused division for embedding, deliberately worsening typical measures of embedding quality to improve the partial solutions we obtain. We apply this approach first to the highly irregular graph of an integer factorisation problem and, passing this initial test, move on to consider more regular geometrically frustrated systems.
Our results show that a problem-focused approach to embedding can improve performance by orders of magnitude.
\end{abstract}

\section{Introduction}
\label{sec:introduction}

Quantum annealing (QA) is a method for energy minimisation that exploits the highly non-classical quantum tunnelling effect. Tunnelling can be a more efficient cooling mechanism than classical thermal fluctuation when the potential landscape contains tall and narrow barriers, making QA a promising near-term heuristic for hard optimisation problems~\cite{hernandez_enhancing_2017,babbush_adiabatic_2014,farhi_quantum_2001,willsch_benchmarking_2022,yaacoby_comparison_2022}. The difficulty of such problems scales exponentially in the number of input bits, so that genuine quantum advantage can be expected with problem sizes well below the 5000 qubit scale of the current D:Wave \textit{Advantage system4.1} device. This device has consequently been a source of much recent interest, with potential applications ranging across machine learning~\cite{benedetti_quantum_2017}, Kosterlitz-Thouless model~\cite{king_observation_2018}, SU(2) lattice gauge theory~\cite{rahman_su2_2021}, financial crisis~\cite{ding_towards_2021}, and portfolio optimisation~\cite{palmer_quantum_2021, phillipson_portfolio_2020}. 

For larger-scale problems the D:Wave quantum annealer rapidly loses its advantage due to the problem of embedding.
For example, when such problems are submitted to the D:Wave quantum hardware, each binary variable may be represented by a chain of physical qubits,
where all physical qubits in such a chain should be in the same state by the end of the annealing process.
Maintaining a chain with many qubits is not easy and the solution may be less frequently optimal due to chain breaks and the \textit{freeze out} phenomenon.
The quantum device therefore displays a problem-hardware trade-off:
Accurate representation of the problem may lead to a less efficient embedding, and consequently lower device performance.

Divide-and-conquer algorithms, splitting problems into smaller pieces for more efficient solution, are well-known in classical computing \cite{rosenberg_graph_2005}. Examples include the sorting algorithms~\cite{sedgewick_implementing_1978} or the Strassen algorithm~\cite{strassen_gaussian_1969} for matrix multiplication.
Divide-and-conquer approaches have also been considered in certain quantum contexts \cite{song_quantum_2022,guerreschi_solving_2021}, and could be helpful to achieve advantage with the D:Wave quantum annealer.

Most existing literature seeking advantage with the D:Wave device relies on minimising the number of physical qubits in the embedding, without regard for the problem structure and in particular for the edge weights \cite{klymko_adiabatic_2014,choi_minor-embedding_2008}.
Here, we demonstrate improved performance with the opposite approach,
dividing the problem graph even at the expense of embedding with a larger physical qubit number.
Partitioning is performed using the Girvin--Newman algorithm for community-detection~\cite{girvin_community_2002,newman_analysis_2004},
and we refer to this approach throughout the text as \textit{Problem-Focused Embedding} (PFE) to contrast it with hardware-focused qubit number minimisation.

As an initial proof of principle, we demonstrate improved performance for problems on irregular graphs where a PFE is most naturally applied.
While we chose integer factorisation~\cite{jiang_quantum_2018} for our demonstration, many other examples such as the anti-ferromagnetic Ising model in scale-free graphs~\cite{bartolozzi_spin_2006} may be cited.
We then look at a class of interesting NP-hard problems on regular structures: Geometrically frustrated lattice systems.
In these systems, ground state degeneracy triggers novel phases of matter including stripe phases~\cite{bramwell_spin_2001,zhao_realization_2020} and spin liquids~\cite{broholm_antiferromagnetic_1990, balents_spin_2010,cepas_heterogeneous_2012,lin_nonmonotonic_2014} characterised by their long-range correlation and the absence of ordinary magnetic order.
In particular, the Kagome lattice provides a prototype of difficult frustrated-lattice problems and can be physically realized in SrCr$_{8-x}$Ga$_{4+x}$O$_{19}$~\cite{ramirez_strong_1990}. Thus, finding the ground state of a Kagome lattice has attracted much attention in condensed matter physics; however, obtaining high-quality solutions using heuristics such as simulated annealing is not easy due to many the local minima and limited gradient information.

Our results show that tight clusters are not required to observe the benefit of a divide-and-conquer strategy. 
As a measure for performance we take $F={-\log_{10}{\left({\epsilon}\right)}}/{\tau}$, where $\tau$ is the mean time required for the quantum device to return each independent set of results and $\epsilon$ is the probability that the device fails to return the global optimum solution (see \cite{king_benchmarking_2015} for a discussion of alternatives).
Despite the fact that a PFE typically requires more qubits (due to the overhead of packing sub-problems in the device), we see better performance metrics.

The remainder of this paper is organised as follows.
In Section~\ref{sec:graph_measures}, we describe problems in quantum annealing in terms of their graph structure, introducing the concepts of sub-graphs and partial solutions.
Our PFE approach is presented, alongside a description of the Girvin--Newman algorithm.
Next, in Section~\ref{sec:experimental_results} experimental results demonstrating advantage in the D:Wave system for both regular and irregular structures are presented. Finally, in Section~\ref{sec:summary_and_conclusions} we summarise our work and discuss ways to identify the potential advantage of QA.

\section{Graph Structure and Quantum Annealing}
\label{sec:graph_measures}

In this section, we discuss the useful graph description of QA problems and introduce the PFE.
We begin with the notion of QA and a limitation arising from embedding. To overcome the limitation, the PFE consists of two following steps: divide the problem graph and then merge the partial solutions.
We shall use the Girvin--Newman algorithm to divide the graph with community structure. We discuss sub-graph reduction and partial solutions using the example of the 1D Ising chain. Based on these concepts, the PFE is introduced using the example of integer factorisation.

\subsection{Viewing Quantum Annealing as a Graph}

QA aims to minimise a quadratic function of binary variables in the form of a ground state search for the Ising-type Hamiltonian~\cite{kadowaki_quantum_1998}
\begin{align}
 	H_{\text{target}}
    &=
    \sum_{i} h_{i} \hat{\sigma}_{z,i}
    +
    \sum_{i>j} J_{ij} \hat{\sigma}_{z,i} \hat{\sigma}_{z,j}
    \,,
    \label{eq:d_wave_system_hamiltonian}
\end{align}
where $\hat{\sigma}_{z,i}$ describes the Pauli-Z operator for the $i$th qubit, while $h_i$ and $J_{ij}$ are the qubit biases and coupling strengths.
This is an NP-Hard problem in general, and its solution is obtained by initialising the system instead in the (known, separable) ground state of the Hamiltonian
\begin{align}
 	H_{\text{initial}}
    &=
    \sum_{i} \hat{\sigma}_{x,i}
    \label{eq:d_wave_initial_hamiltonian}
\end{align}
and then varying the relative magnitudes of $H_{\text{initial}}$ and $H_{\text{target}}$ according to
\begin{align}
    H\left(t\right)
    &=
    f\left(t\right)
 	H_{\text{initial}}
    +
    g\left(t\right)
    H_{\text{target}},
    \nonumber\\
    g\left(0\right) &= 0,
    \nonumber\\
    \lim_{t\rightarrow\infty}f\left(t\right) &= 0
    \,,
    \label{eq:d_wave_time_dependent_hamiltonian}
\end{align}
where $f\left(t\right)$ and $g\left(t\right)$ are real, non-negative, monotonic functions of time $t$.
$f\left(t\right)$ is said to be the strength of \textit{quantum fluctuations}, as the $\hat{\sigma}_{x}$ terms in $H_{\text{initial}}$ facilitate transitions between the eigenstates of $H_{\text{target}}$.
In the adiabatic limit, the target ground state is obtained deterministically.
Due to noise and time constraints, adiabatic evolution is rarely observed in practice;
quantum tunnelling becomes the key mechanism for computational advantage.

QA is often compared to digital simulated annealing (SA), where the temperature parameter plays a similar role to the strength of quantum fluctuations. 
In SA, the idea is to use thermal fluctuations to allow the system to escape from local minima of the potential landscape so that the system reaches the global minimum under an appropriate annealing schedule (rate of decrease of temperature). For an energy barrier $\Delta E$, the escape probability is $\sim\exp(-\Delta E/T)$ at temperature $T$. In QA, the system will likely tunnel through the barrier if it is narrow: The tunnelling probability is $\sim\exp(-w\sqrt{\Delta E}/\Gamma)$ for tunnelling fluctuation field $\Gamma$ and barrier width $w$. For high and narrow potential barriers, QA utilising the tunnelling effect can therefore be more efficient.

Connectivity constraints do not affect $H_{\text{initial}}$, as it contains no coupling terms.
The coupling terms in $H_{\text{target}}$ must however be mapped to interactions in the physical device.
A coupling term in $H_{\text{target}}$ may be simulated either in the form of direct interactions between neighbouring physical qubits, or indirectly in the form of a chain of interacting ancillae.
Considering these qubits as nodes of a graph, and coupling terms or interactions as its edges, the problem of mapping an optimisation problem to the quantum annealing device corresponds to the graph problem of \textit{minor embedding}:
A first graph is said to be a \textit{minor} of a second if it can be obtained by deleting vertices and contracting edges (identifying neighbouring vertices) in this second graph.
A quadratic unconstrained binary optimisation (QUBO) problem can be embedded in a quantum annealing device if the graph of coupling terms in $H_{\text{target}}$ can be expressed as a minor of the graph of physical qubit interactions in the device.
The typical measure of the quality of such an embedding is the overhead in the number of physical qubits, which corresponds to the number of edge-contractions required to obtain the logical coupling graph from the physical interaction graph.

\subsection{Sub-graphs and Partial Solutions}

\begin{figure*}[t]
    \centering
    \includegraphics[width=1.00\textwidth]{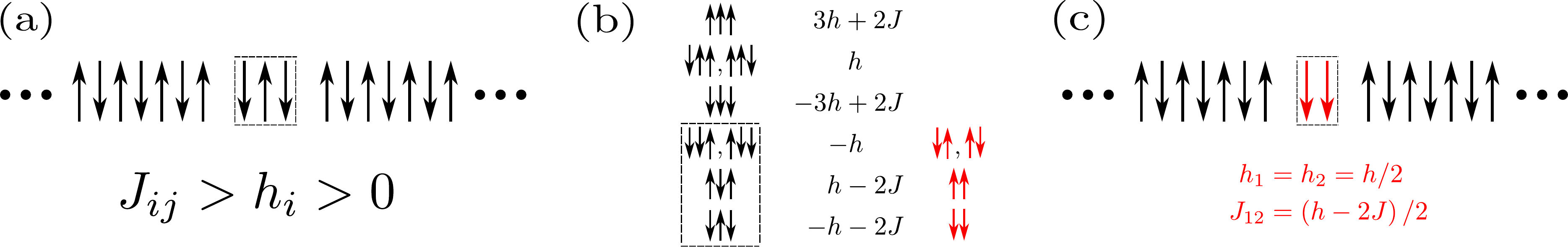}
    \caption{\label{fig:1d_ising_chain_subgraph_reduction}
        Example of sub-graph reduction for the 1D Ising chain.
        Here $J_{ij}$ terms along the $1$D chain are all equal initially, as are $h_{i}$ terms.
        (a)
        The initial Ising chain. Spins have been arranged at a global optimum to emphasise how this is preserved, though the specific configuration of the chain will not be known initially.
        Spins comprising the sub-graph to be reduced are boxed and separated in the centre.
        (b)
        Enumeration of configurations and energies for the three spins in the sub-graph to be reduced, and identification (in red) of the four lowest-energy solutions exhausting all possible boundary-spin configurations.
        (c)
        The modified Ising chain. The modified red component of the graph is boxed and separated in the centre. Also shown in red are the modified self-energy and interaction terms for the two spins in this component. These new terms are chosen to preserve the energy level structure associated with the low-energy configurations of the original sub-graph.
    }
\end{figure*}

For many problems, we can identify ahead of time that the solution is contained in only a small region of the total configuration space.
Consider the $1$D periodic chain as a simple example, as displayed in
Fig.~\ref{fig:1d_ising_chain_subgraph_reduction}.
We begin by looking at only a small connected segment, enumerating all possible solutions and their energies, which we will refer to as the segment's `internal configurations'.
Now, when this segment is incorporated into the global solution for the total chain, the appropriate internal configuration will depend only on the states of those bits on the boundary between the segment and the rest of the chain --- all other bits within the segment are used only to select the internal configuration of minimal energy satisfying these boundary conditions.
This means that we can ignore all but four solutions from the internal segment, taking the minimal-energy solution for each of the four boundary-bit configurations.
The problem may then be remapped so that the initial segment is represented by only the two boundary bits, with energies and an interaction chosen to correspond to the energy gaps for the four solutions we retain.
Applying this method recursively, the $1$D chain (or any tree structure) can be solved in linear time regardless of the problem parameters.

Remapping the problem to eliminate variables comes at no cost when the problem graph is a tree structure.
This may be highly significant for scale-free graphs and others with many nodes of degree one or two.
However, for many problems we expect to be presented with a much more densely connected graph, with many small but non-zero edges corresponding to measurement samples.
For these more general graphs such elimination introduces new interaction terms of order up to the degree of the removed node \cite{guerreschi_solving_2021}.
Conversely, reducing the order of interactions (or `quadratization') requires the introduction of new variables \cite{dattani_quadratization_2019}.
While divide-and-conquer strategies have been applied in the more general context of quantum approximate optimization algorithms (QAOA) \cite{li_large-scale_2022}, they have thus far been capable only of limited application in QUBO-restricted experimental quantum annealing.
We next introduce a modified divide-and-conquer method suitable for quantum annealing.

\subsection{A Divide-and-Conquer Method for QUBO-Restricted Quantum Annealing}
\label{sub:a_divide_and_conquer_method_for_qubo_restricted_quantum_annealing}

\begin{figure*}[t]
\begin{center}
    \includegraphics[width=0.8\textwidth]{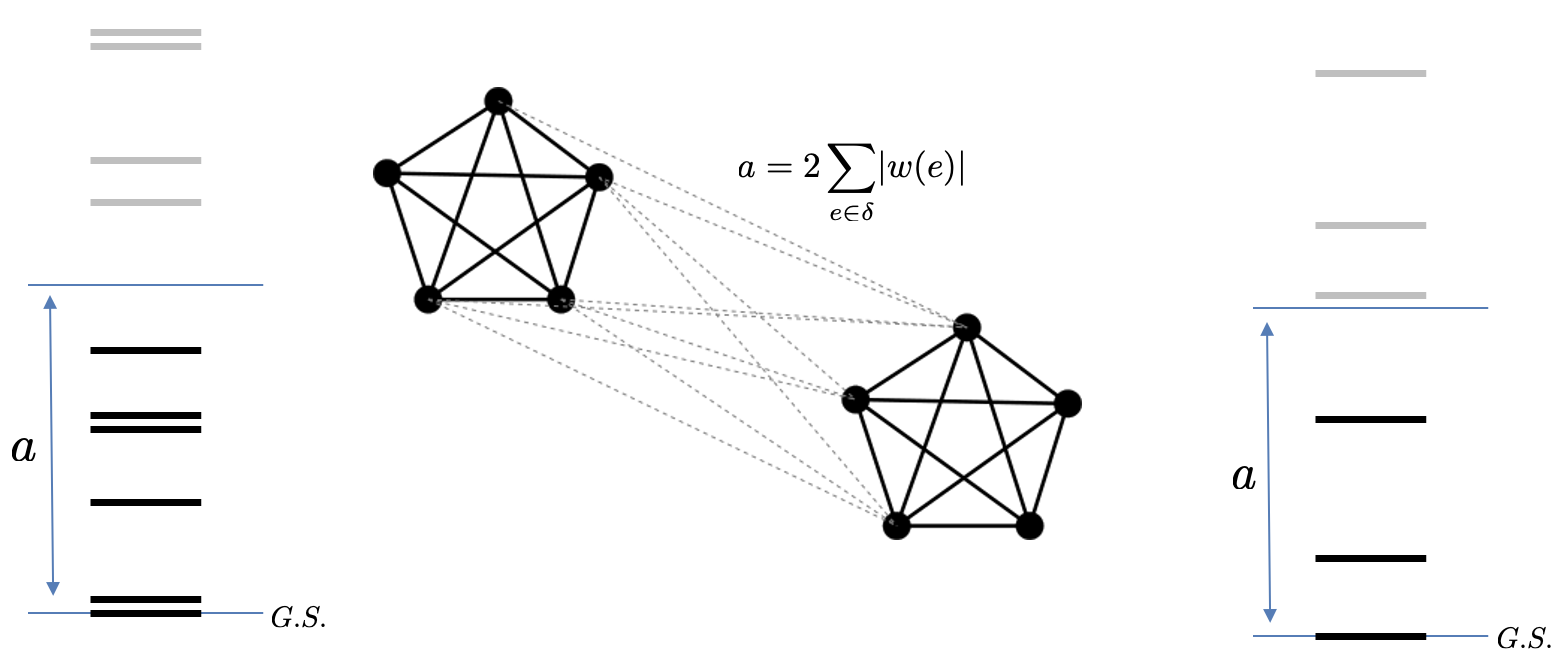}
    \caption{\label{fig:cluster_subsample}
        Diagrammatic representation of sampling local solutions in a tightly clustered problem graph.
        Here the problem graph is grouped into two tightly connected clusters, with a number of weaker edges between the two clusters.
        The value $a$, twice the sum of boundary edge magnitudes, determines an upper bound on the energies of local configurations.
        Configurations exceeding this bound may be discarded, reducing the local search space.
        Supposing $N$ valid local configurations exist in the first cluster, and $M$ in the second,
        the overall problem complexity is then reduced to the sum of the individual sub-problems, and an additional merging term of order $N\cdot M$.
    }
\end{center}
\end{figure*}

It is intuitive that solutions for distinct clusters should retain their form in spite of interconnecting edges, when those edges have very small weights.
This intuition is supported by the fact that the eigenvalues $\lambda_{k}$ of an Hermitian matrix $A$ are
well-conditioned under small perturbations $\delta A$,
\begin{align}
    \max_{k}
    \left\lvert
        \lambda_{k} \left( A + \delta A \right)
        -
        \lambda_{k} \left( A \right)
    \right\rvert
    \leq
    \left\lVert
        \delta A
    \right\rVert_{2}
    ,
\end{align}
and is implicit in spectral methods for graph sparsification \cite{batson_spectral_2013}.
Consider two highly connected clusters connected by a number of weak edges, as represented diagrammatically in Figure~\ref{fig:cluster_subsample}.
While the ground state of each cluster may no longer correspond to the target solution for the entire graph, we observe that the energy penalty introduced by the connecting weak edges is bounded above by a value equal to twice the sum of the edge weight magnitudes (\textit{twice} because variables take the values $\pm1$).
Call this bound $a$.
When the global solution is projected onto a local state in each cluster, each resulting local state has an energy at most $a$ larger than its local ground state; solutions with local energies exceeding this bound could be improved by simply choosing the local ground state, regardless of the boundary configuration.
Supposing that $N$ of the total $2^{S}$ internal energies fall within the gap $a$ of the ground state ($S$ nodes in the cluster), the cluster is then effectively reduced to the smaller $N$-dimensional system.
In fact, once these low energy solutions have been determined, a stricter gap follows using a boundary sum with edges from only those nodes where each solution differs from the local ground state.

One remaining question is how to sample efficiently from our restricted set of local solutions, which may differ widely from one another in their particular configurations.
Fortunately, unlike other heuristics such as genetic or Tabu search \cite{boros_local_2007,kochenberger_unconstrained_2014}, simulated and quantum annealing are capable of sampling states with a wider range of energies in the global configuration space by manipulating the cooling schedule and final temperature.

\subsection{A Problem-Focused Embedding}
\label{sub:a_problem_focused_embedding}

\begin{figure*}[!ht]
    \centering
    \includegraphics[width=0.85\textwidth]{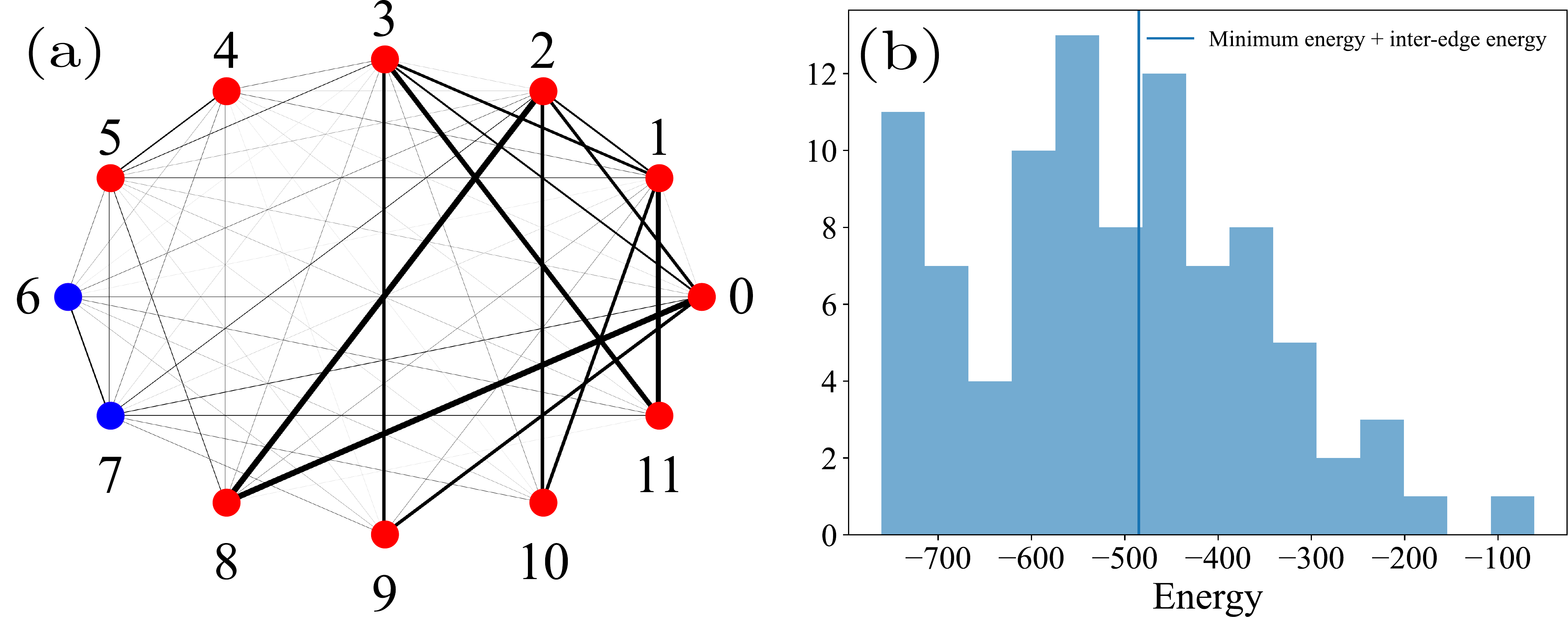}
    \caption{\label{fig:pfe_example}
        Example of the PFE for integer factoring problem $143=13\times 11$ with 12 qubits.
        (a) The problem graph is partitioned as two sub-graphs (the ten red nodes and two blue nodes) by the Girvin--Newman algorithm.
        The thickness of the line represents the edge weight.
        (b) The sampled energy distribution for the larger red sub-graph.
        Only configurations lower than the blue boundary line are considered in searching for global solutions.
        The final temperature (quantum fluctuation magnitude) has been chosen to sample this region of interest.
        }
\end{figure*}

To implement our divide-and-conquer strategy, we use a simple partitioning method based on the edge \textit{betweenness centrality}, defined as the number of the shortest paths that go through an edge in a graph \cite{freeman_set_1977}.
The problem graph is divided into sub-graphs using the weighted Girvin--Newman algorithm~\cite{girvin_community_2002,newman_analysis_2004}.
This algorithm detects communities from the problem graph by gradually removing bottleneck edges between them.
When several edges have the same betweenness centrality, one is removed at random.

Figure~\ref{fig:pfe_example} shows a simple application of our divide-and-conquer strategy to the factorization of $143=13\times 11$.
In the QUBO representation, this problem reduces to a graph with 12 nodes.
The problem graph is divided into smaller sub-graphs, each sub-graph is solved by QA or SA, and we merge the solutions.
Hereafter, we focus only on bipartition, as a proof of concept.

\section{Experimental results}
\label{sec:experimental_results}

Next we apply the PFE of Section~\ref{sec:graph_measures} to several small-scale examples of NP-hard problems.
This approach is naturally adopted for irregular problem graphs such as that of the integer factoring problem described in Fig.~\ref{fig:pfe_example}, and so we begin there.
We then apply the PFE approach to regular, geometrically-frustrated systems, to show that tight clusters are not always required.

Our results $F/F_{SA}$ are expressed relative to a standard application of digital simulated annealing\footnote{using a single process on an Apple M1 chip}. This unit of measure is arbitrary and we note that, particularly for irregular graph structures, SA does not necessarily represent the most efficient classical method.
We choose this form both for succinctness in the tables to follow and to emphasise the increase quantum advantage, as distinct from any bare classical advantage arising from our division of the problem.

\subsection{An Irregular Graph Structure}

Problem graphs with irregular structures introduce two complications:
\begin{enumerate}
    \item
    They have broad structural features, often exploitable via specific classical strategies.
    \item
    They are highly non-planar and usually contain long-range interactions, making device embedding difficult.
\end{enumerate}
These constraints imply the need for hybrid classical and quantum solutions.

One application that generates irregular problem graph structures is integer factorisation.
Integer factorisation is an NP-hard problem for a classical computer, and consequently forms the basis of RSA public key cryptography. 
While Shor’s algorithm achieves polynomial complexity for a hypothetical universal quantum computer, the largest number factored using Shor's algorithm at present is 21~\cite{martin_experimental_2012}, so there is still a long way to go before practically meaningful factorisation is achieved in this manner.
Recently, QA with the D:Wave device was proposed as a factorisation heuristic using a transformation that maps the problem into a QUBO model~\cite{jiang_quantum_2018}.

Table~\ref{tab:pfe_factoring} shows the relative performance $F/F_{\rm SA}$ of standard QA and the PFE for the factorisation of $1591=43\times 37$.
\begin{table*}[!ht]
    \begin{center}
        \caption{Results of the relative performance metric of factoring $1591=43\times 37$ with 29 qubits using the standard and PFE approaches.
        QA is performed by D:Wave \textit{Advantage\_system4.1}.
        The number of samples is $10^3$.
        The number of physical nodes $N_{\rm P}$, the sample estimate for the success probability using Bayes' inference $p_{\rm Bayes}$, the total run time in microseconds (including overheads) $\tau_{\rm Real}$, and performance metric $F/F_{\rm SA}$ (relative to that of a standard application of SA) are listed.
        }
        \vskip 1mm
        \setlength{\tabcolsep}{15pt}
        {\renewcommand{\arraystretch}{1.5}
            \begin{tabular}{ccccc}
                \hline
                \hline
                Method
                & $N_{\rm P}$ 
                & $p_{\rm Bayes}$
                & $\tau_{\rm Real}$
                & $F/F_{\rm SA}$\\
                \hline
                Standard QA
                & $73$
                & $10^{-3}$
                & $170,302$
                & $0.26$\\
                PFE (QA$+$QA)
                & $66$ \& $2$
                & $0.17$
                & $405,995$
                & $19.98$\\
                PFE (SA$+$SA)
                & $-$
                & $0.17$
                & $652,691$
                & $12.43$\\
                \hline
                \hline
        \end{tabular}}
        \label{tab:pfe_factoring}
    \end{center}
\end{table*}
The relative performance metric of the PFE is enhanced approximately $76$ times beyond standard QA (as proposed, for instance, in \cite{jiang_quantum_2018}) and the required number of physical qubits is decreased.
Applying the same divide-and-conquer strategy in an entirely classical manner to SA, a relative improvement of only about $12$ times is observed over the standard approach.

\subsection{A Regular Planar Structure}

The regular structures of geometrically frustrated lattices in many-body physics have two attractive qualities for QA:
\begin{enumerate}
    \item
    Their low degree and local interactions should ease embedding in the quantum device.
    \item
    The lack of clustering or other broad structure means that heuristics such as simulated annealing may often be the most fruitful approach.
\end{enumerate}
Despite these advantages, the performance of standard QA rapidly decreases as we near about one hundred qubits, due to growing embedding constraints.
This rough $100$ qubit limit presents a challenge for the practical application of QA to many body physics.

\begin{figure}[!ht]
\begin{center}
    \includegraphics[width=0.35\textwidth]{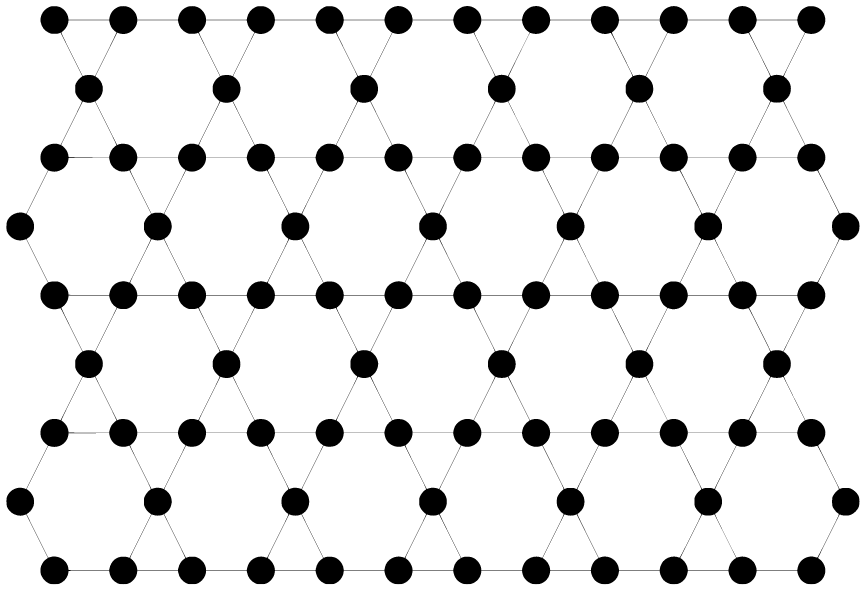}
    \caption{\label{fig:kagome_black_and_white}
        The (periodic) Kagome lattice, or trihexagonal tiling, forms a degree-$4$ regular graph, and may be obtained by deleting nodes in a triangular lattice.
    }
\end{center}
\end{figure}

A divide-and-conquer approach may help to raise the tolerable qubit number.
It is not immediately clear, however, whether we should expect to observe advantage in problem division for a highly regular structure.
Next we therefore investigate the performance of a PFE approach to ground state search on the Kagome lattice, shown in Figure~\ref{fig:kagome_black_and_white}.
The Kagome lattice is a prototypical example of geometric frustration in many body physics.

Table~\ref{tab:pfe_kagome} shows the relative performance $F/F_{\rm SA}$ of standard QA and the PFE approach for QA and SA.
\begin{table*}[!ht]
    \begin{center}
        \caption{Results of the relative performance metric of the Kagome lattice with $90$ qubits using the standard QA and the PFE approach to both QA and SA.
        QA is performed by D:Wave \textit{Advantage\_system4.1}.
        The number of samples is $10^3$ and $h=J$.
        The number of physical nodes $N_{\rm P}$, the sample estimate for the success probability using Bayes' inference $p_{\rm Bayes}$, the total run time in microseconds (including overheads) $\tau_{\rm Real}$, and performance metric $F/F_{\rm SA}$ (relative to that of a standard application of SA) are listed.}
        \vskip 1mm
        \setlength{\tabcolsep}{15pt}
        {\renewcommand{\arraystretch}{1.5}
            \begin{tabular}{ccccc}
                \hline
                \hline
                Method
                & $N_{\rm P}$ 
                & $p_{\rm Bayes}$
                & $\tau_{\rm Real}$
                & $F/F_{\rm SA}$\\
                \hline
                Standard QA
                & $127$
                & $0.02$
                & $152,904$
                & $2.30$\\
                PFE (QA$+$QA)
                & $54$ \& $48$
                & $0.92$
                & $527,265$
                & $69.47$\\
                PFE (SA$+$SA)
                & $-$
                & $0.92$
                & $1,537,878$
                & $23.83$\\
                \hline
                \hline
        \end{tabular}}
        \label{tab:pfe_kagome}
    \end{center}
\end{table*}
In comparison to a standard application of QA or SA, the PFE approach increases the total run time due to the final matching of component solutions, but increases to an even greater degree the probability of discovering the global optimum with each sample.
We observe that the PFE approach improves the performance by approximately $30$ times for QA ($69/2.3$), and approximately $24$ times when applied classically for SA.
While the bulk of the improvement seems therefore classical, there is nonetheless a significant further boost that is unique to the quantum device.

\section{Discussion}
\label{sec:summary_and_conclusions}

We have proposed and tested experimentally a problem-focused divide-and-conquer approach to quantum annealing.
Dividing the problem graph improves the quality of partial solutions by improving the embedding quality for interactions within each partition, facilitating quantum advantage.
We successfully applied this approach to irregular graph structures, where advantage is most expected, but also to highly regular geometrically frustrated lattices.

We expect that the improved performance we observe for frustrated lattices with many qubits will be applicable in many-body physics:
Frustration in the Kagome lattice can lead to the emergence of a flat band, with strongly-correlated atomic behaviour determined by interactions and topology. Our initial results suggest that QA with the D:Wave system could be fruitfully applied to ground state discovery in strongly-correlated systems, identifying critical phenomena in frustrated models and predicting experimental results in strongly-correlated materials.

In contrast to similar recent work \cite{guerreschi_solving_2021}, our approach requires only quadratic interaction terms, and is therefore compatible with D:Wave quantum devices.
Further, the residual problem complexity depends in our case on the ratio between boundary edge weights and the spectral density of low-energy states, rather than the boundary node number.

For our proof-of-concept results in Section~\ref{sec:experimental_results}, we implemented community detection using the weighted Girvin--Newman algorithm. We acknowledge however that, community detection being a well studied problem, alternative methods for community detection may improve performance.
We use a performance metric making use of knowledge of the global minimum (through $\epsilon$, the probability of failing to return this minimum value).
Presently, D:Wave devices demonstrate quantum advantage with problems roughly on the order of $150$ qubits.
We expect therefore that problems solved on contemporary quantum annealing devices will not be those that cannot be solved by a classical machine, but rather those for which a classical machine may solve the problem in a less cost-efficient manner.
As a result, the use of global information for test problems of comparable size is justified.

We identify two interesting questions arising from our results, which could motivate future work:
Firstly, though we restrict ourselves to a bipartition for proof-of-concept, three or perhaps multiple sub-graphs could become optimal as the problem size increases.
How might the density of low-energy states be estimated to assess the trade-off between a more efficient embedding as a result of further sub-division and the cost of merging partial solutions?
Secondly, rather than solving each partition separately, it may prove fruitful to develop a whole-problem embedding, giving priority to intra-community links.
At present, strong coupling in ancillary chain structures interferes with the solution in each partition.
Is an alternative construction for ancillary chains possible, that would allow communities to interact coherently without such strong interference?

\section*{Acknowledgments}
This research was supported by the education and training program of the Quantum Information Research Support Center, funded through the National research foundation of Korea (NRF) by the Ministry of science and ICT (MSIT) of the Korean government, No.~2021M3H3A103657313 (MJ).
We acknowledge the Samsung GRC project and the UK Hub in Quantum Computing and Simulation, part of the UK National Quantum Technologies Programme with funding from UKRI EPSRC grant EP/T001062/1.
This work was supported by the Engineering and Physical Sciences Research Council of the UK (EPSRC) Grant number EP/W032643/1.

\section*{References}
\bibliographystyle{iopart-num}
\bibliography{bibliography}

\end{document}